\documentclass[aps,pra,reprint,groupedaddress,floatfix]{revtex4-2} 

\usepackage{amsmath}
\usepackage{amssymb}
\usepackage{graphicx} 
\usepackage{braket} 
\usepackage[colorlinks=true,linkcolor=blue,citecolor=blue,urlcolor=blue]{hyperref} 
\usepackage{siunitx} 
\usepackage{microtype} 

\begin{document}

\title{Impact of Temporally Correlated Dephasing Noise on the Fidelity of the 2-Qubit Deutsch-Jozsa Algorithm}

\author{Souvik Ghosh} 
\email{souvikghosh2012@gmail.com} 
\affiliation{Adamas University, Kolkata, West Bengal, India} 

\date{\today}

\begin{abstract}
Understanding the influence of realistic noise on quantum algorithms is paramount for the advancement of quantum computation. While often modeled as Markovian, environmental noise in quantum systems frequently exhibits temporal correlations, leading to non-Markovian dynamics that can significantly alter algorithmic performance. This paper investigates the impact of temporally correlated dephasing noise, modeled by the Ornstein-Uhlenbeck (OU) process, on the fidelity of the 2-qubit Deutsch-Jozsa algorithm. We perform numerical simulations using Qiskit, systematically varying the noise strength ($\sigma_{\text{OU}}$) and correlation time ($\tau_c$) of the OU process. Our results demonstrate that the algorithm's fidelity exhibits a non-monotonic dependence on $\tau_c$, particularly at higher noise strengths, with certain intermediate correlation times proving more detrimental than others. We find that a standard Markovian dephasing model, matched to the single-step error variance of the OU process, accurately predicts fidelity only in the limit of very short correlation times. For longer correlation times, the Markovian approximation often overestimates the algorithm's fidelity, failing to capture the complex error dynamics introduced by the noise memory. These findings highlight the necessity of incorporating non-Markovian characteristics for accurate performance assessment of quantum algorithms on near-term devices and underscore the limitations of simpler, memoryless noise models.
\end{abstract}

\maketitle

\section{Introduction}
\label{sec:introduction}

Quantum computing holds the promise of revolutionizing computation by harnessing the principles of quantum mechanics to solve problems currently intractable for even the most powerful classical computers \cite{NielsenChuangBook, PreskillNotes}. However, the realization of fault-tolerant quantum computers faces significant hurdles, chief among them being the decoherence of quantum states due to unavoidable interactions with their environment \cite{Zurek2003_Decoherence}. This interaction, broadly termed "noise," corrupts quantum information and limits the fidelity of quantum operations, posing a major challenge for the development of Noisy Intermediate-Scale Quantum (NISQ) devices and the path towards scalable quantum computation \cite{Preskill2018_NISQ}.

Among the various types of quantum noise, dephasing, or phase damping, is particularly detrimental as it destroys the phase coherence essential for many quantum algorithms that rely on interference phenomena. Traditionally, noise in quantum systems is often modeled under the Markovian approximation, which assumes that the environment has no memory and that errors occurring at different times are uncorrelated \cite{BreuerPetruccioneBook}. While mathematically convenient, this approximation may not accurately capture the dynamics of many realistic quantum systems where environmental fluctuations can exhibit temporal correlations, leading to non-Markovian noise effects \cite{deVega2017_ReviewNonMarkovian, Lidar2019_ReviewNonMarkovian}. Such correlations mean that the noise "remembers" its past, influencing future error probabilities in a complex manner. Understanding the impact of these non-Markovian characteristics is crucial for accurately predicting qubit performance and for developing robust quantum error correction and mitigation strategies \cite{Terhal2015_QECReview}.

The Deutsch-Jozsa algorithm \cite{Deutsch1992, Cleve1998_DJ} serves as a paradigmatic example of quantum computational speedup, demonstrating how a quantum computer can determine a global property of a function (whether it is constant or balanced) with a single query to a quantum oracle, whereas a classical deterministic algorithm would require multiple queries. Due to its relative simplicity and its clear reliance on quantum interference, the Deutsch-Jozsa algorithm provides an excellent testbed for investigating the effects of different noise models on fundamental quantum computational tasks. While the impact of Markovian noise on various quantum algorithms has been extensively studied, the specific effects of temporally correlated dephasing noise on the Deutsch-Jozsa algorithm, particularly how varying noise correlation times influence its fidelity, have received less focused attention. This represents an important gap, as real-world noise on quantum hardware often possesses non-trivial temporal structures \cite{Martinis2003_LowFreqNoise, Bylander2011_1fNoise}.

In this paper, we numerically investigate the impact of temporally correlated dephasing noise, modeled as an Ornstein-Uhlenbeck (OU) process \cite{Uhlenbeck1930}, on the fidelity of the 2-qubit Deutsch-Jozsa algorithm. We systematically explore how the algorithm's success probability is affected by varying the key parameters of the OU noise: its strength ($\sigma_{\text{OU}}$) and, critically, its correlation time ($\tau_c$). By comparing these results with simulations under a standard Markovian dephasing model of equivalent strength, we aim to elucidate the distinct signatures of non-Markovian dynamics and identify regimes where the Markovian approximation proves insufficient.

This paper is structured as follows. In Section~\ref{sec:methods}, we detail the Deutsch-Jozsa algorithm, describe the OU noise model and its application to the quantum circuit, and outline our simulation methodology. Section~\ref{sec:results} presents the core simulation findings, focusing on the fidelity as a function of noise strength and correlation time. In Section~\ref{sec:discussion}, we interpret these results, discussing the implications of non-Markovian effects and the limitations of Markovian approximations. Finally, Section~\ref{sec:conclusion} summarizes our main conclusions and provides an outlook for future research.

\section{Methods}
\label{sec:methods}

\subsection{The 2-Qubit Deutsch-Jozsa Algorithm}
The Deutsch-Jozsa (DJ) algorithm distinguishes between constant and balanced Boolean functions $f: \{0,1\}^n \to \{0,1\}$ using a quantum oracle \cite{Deutsch1992}. In this work, we focus on the 2-qubit version ($n=1$ for the query qubit, plus one ancilla qubit). The query qubit (qubit 0) is initialized to $\ket{+} = (\ket{0} + \ket{1})/\sqrt{2}$ by applying a Hadamard (H) gate to $\ket{0}$. The ancilla qubit (qubit 1) is initialized to $\ket{-} = (\ket{0} - \ket{1})/\sqrt{2}$ by applying an X gate then an H gate to $\ket{0}$.
The oracle $U_f$ implements the transformation $\ket{x}\ket{y} \to \ket{x}\ket{y \oplus f(x)}$. For the 2-qubit case, this effectively imprints the phase $(-1)^{f(x)}$ onto the query qubit when the ancilla is in $\ket{-}$: $U_f \ket{x}\ket{-} = (-1)^{f(x)}\ket{x}\ket{-}$.
After the oracle, an H gate is applied to the query qubit. Measurement of the query qubit in the computational basis then deterministically yields $\ket{0}$ if $f(x)$ is constant, or $\ket{1}$ if $f(x)$ is balanced. Our simulations primarily utilized a balanced oracle where $f(x)=x$ (implemented via a CNOT gate with qubit 0 as control and qubit 1 as target).

\subsection{Temporally Correlated Dephasing Noise Model}
To model temporally correlated dephasing noise, we employed the Ornstein-Uhlenbeck (OU) process \cite{Uhlenbeck1930, GardinerBook}. The OU process $\nu(t)$ describes a Gaussian, stationary, and Markovian random process whose stochastic differential equation (SDE) is given by:
\begin{equation}
    d\nu(t) = -\frac{1}{\tau_c} \nu(t) dt + \sqrt{\frac{2\sigma_{\text{OU}}^2}{\tau_c}} dW(t)
    \label{eq:ou_sde_paper}
\end{equation}
where $\tau_c$ is the correlation time of the noise, $\sigma_{\text{OU}}^2$ is the stationary variance of $\nu(t)$ (characterizing the noise strength), and $dW(t)$ is the increment of a Wiener process. The process $\nu(t)$ represents fluctuations in, for example, a background field or control parameter, leading to phase errors on the qubit.

For numerical simulation, the OU process was discretized using the exact update formula for time steps $\Delta t_{\text{step}}$:
\begin{equation}
    \nu_{k+1} = \nu_k e^{-\Delta t_{\text{step}}/\tau_c} + \sigma_{\text{OU}} \sqrt{1 - e^{-2\Delta t_{\text{step}}/\tau_c}} N_k(0,1)
    \label{eq:ou_discrete_paper}
\end{equation}
where $\nu_k = \nu(k \Delta t_{\text{step}})$ and $N_k(0,1)$ is a random number drawn from a standard normal distribution at each step $k$. The initial value $\nu_0$ was drawn from the stationary distribution $\mathcal{N}(0, \sigma_{\text{OU}}^2)$.
The dephasing error experienced by the qubit at step $k$ was modeled as an accumulated phase $\phi_k$, proportional to the OU process value $\nu_k$ and the duration of the step:
\begin{equation}
    \phi_k = \nu_k \Delta t_{\text{step}}
    \label{eq:phase_error_paper}
\end{equation}
In our simulations, we set $\Delta t_{\text{step}} = 0.1$ (arbitrary time units, with $\tau_c$ expressed as a multiple of this unit).

\subsection{Noise Application in Quantum Circuit Simulations}
All quantum circuit simulations were performed using Qiskit \cite{Qiskit}. The temporally correlated dephasing noise was applied exclusively to the query qubit (qubit 0). The noise was introduced as discrete phase error operations at $N_{\text{noise\_points}} = 3$ specific points within the algorithm's execution:
\begin{enumerate}
    \item After the initial Hadamard gate on the query qubit.
    \item After the oracle $U_f$ operation.
    \item After the final Hadamard gate on the query qubit.
\end{enumerate}
At each such point $j \in \{1, 2, 3\}$, a phase rotation gate $R_z(\phi_j) = \text{diag}(1, e^{i\phi_j})$ was applied, where $\phi_j$ was the $j$-th value obtained from a single realization of the OU noise sequence $\{\nu_k\}$ generated as per Eq.~\eqref{eq:ou_discrete_paper} and scaled by Eq.~\eqref{eq:phase_error_paper}. Each sequence $\{\phi_j\}$ used for one noisy circuit execution was drawn from a distinct OU trajectory.

\subsection{Markovian Dephasing Model for Comparison}
For comparison, we also simulated the DJ algorithm under a standard Markovian dephasing model. This was implemented using Qiskit's \texttt{phase\_damping\_error}, which describes a single-qubit phase damping channel. The Kraus operators for this channel with damping parameter $\lambda_{\text{pd}}$ are $E_0 = \text{diag}(1, \sqrt{1-\lambda_{\text{pd}}})$ and $E_1 = \text{diag}(0, \sqrt{\lambda_{\text{pd}}})$. This channel causes the off-diagonal elements $\rho_{01}$ of the density matrix to decay as $\rho_{01} \to \sqrt{1-\lambda_{\text{pd}}} \rho_{01}$ per application.
To establish an equitable comparison between the OU and Markovian models, the damping parameter $\lambda_{\text{pd}}$ was related to the variance of the phase error per step from the OU model. For small phase errors $\phi_k$, the average effect of the OU noise on coherence is $\langle e^{i\phi_k} \rangle \approx e^{-\text{Var}(\phi_k)/2} \approx 1 - \text{Var}(\phi_k)/2$. Equating this to $\sqrt{1-\lambda_{\text{pd}}} \approx 1 - \lambda_{\text{pd}}/2$ (for small $\lambda_{\text{pd}}$), we set:
\begin{equation}
    \lambda_{\text{pd}} \approx \text{Var}(\phi_k) = (\Delta t_{\text{step}} \sigma_{\text{OU}})^2
    \label{eq:lambda_pd_matching}
\end{equation}
This Markovian dephasing error was applied to the query qubit at the same $N_{\text{noise\_points}}=3$ locations within the circuit as the OU noise.

\subsection{Simulation Parameters and Fidelity Metric}
The impact of correlated noise was investigated by varying the OU noise strength parameter $\sigma_{\text{OU}}$ over the range $[1.0, 5.0]$ (in steps of \SI{1.0}{}, corresponding to units of rad/$\Delta t_{\text{step}}$ for $\nu_k$) and the correlation time $\tau_c$ over the range $[0.1, 10.0]$ (in units of $\Delta t_{\text{step}}$), using the specific values $\{0.1, 0.5, 1.0, 2.0, 5.0, 10.0\}$. Some results also incorporate a finer scan of $\tau_c$ values for $\sigma_{\text{OU}} \in \{4.0, 5.0\}$, including $\{1.5, 2.5, 3.0, 4.0, 6.0, 7.0, 8.0\}$ in addition to the base set.

For each pair of $(\sigma_{\text{OU}}, \tau_c)$ parameters, the fidelity of the DJ algorithm was determined by averaging the success probability over $N_{\text{traj}} = 100$ independent OU noise trajectories. For each trajectory (i.e., each specific noisy circuit realization), the success probability was estimated from $N_{\text{shots}} = 1024$ simulated measurements (shots) of the query qubit. For the Markovian simulations, results were obtained from $N_{\text{shots}} \times N_{\text{traj}}$ total shots for comparable statistical weight.
The fidelity (success probability) was defined as the probability of measuring the query qubit in the state $\ket{1}$ for the balanced oracle $f(x)=x$.

\section{Results}
\label{sec:results}

We investigated the fidelity of the 2-qubit Deutsch-Jozsa algorithm, configured with a balanced oracle ($f(x)=x$), under the influence of temporally correlated dephasing noise modeled by an Ornstein-Uhlenbeck (OU) process. The noise was applied to the query qubit at three distinct points during its evolution. The simulations were conducted for a range of OU noise strengths ($\sigma_{\text{OU}}$) and correlation times ($\tau_c$), with results averaged over 100 noise trajectories and 1024 measurement shots per trajectory. For comparison, simulations under a Markovian dephasing model with an equivalently matched single-step error variance were also performed.

\subsection{Impact of Noise Correlation Time on Fidelity}
\begin{figure}[htbp]
  \centering
  \includegraphics[width=0.95\columnwidth]{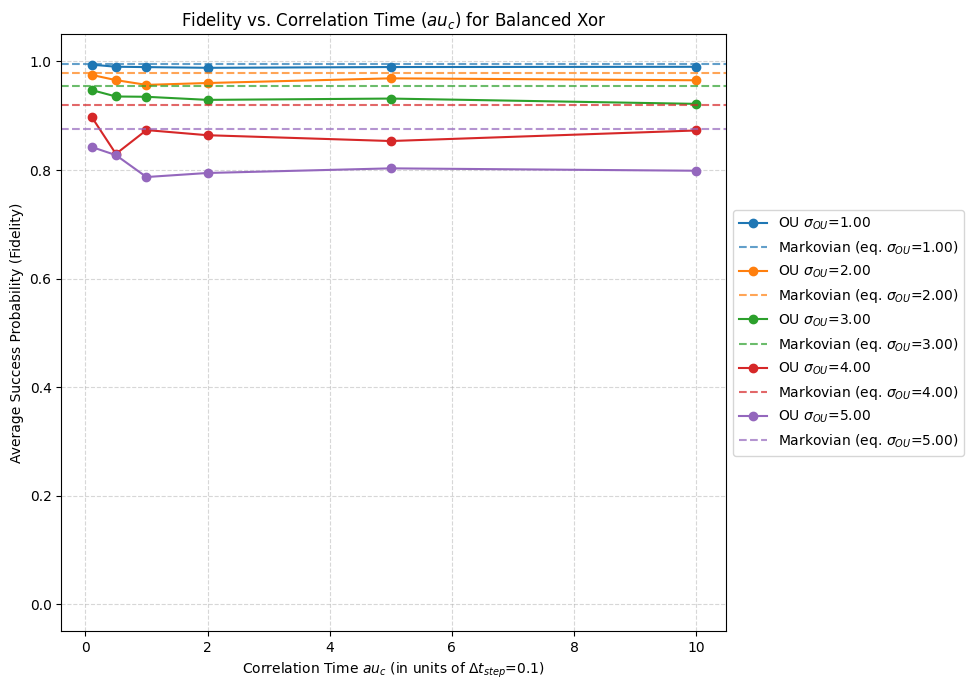}
  \caption{Fidelity of the 2-qubit Deutsch-Jozsa algorithm (balanced oracle) as a function of noise correlation time $\tau_c$ (in units of $\Delta t_{\text{step}}=0.1$) for various OU noise strengths $\sigma_{\text{OU}}$. Dashed lines represent the fidelity under an equivalent Markovian dephasing model. Results averaged over 100 noise trajectories.}
  \label{fig:fid_vs_tau_c_full_sigma}
\end{figure}

The dependence of the algorithm's fidelity on the noise correlation time $\tau_c$ (expressed in units of $\Delta t_{\text{step}}=0.1$) for different noise strengths $\sigma_{\text{OU}}$ is presented in Fig.~\ref{fig:fid_vs_tau_c_full_sigma}.

For relatively low noise strength ($\sigma_{\text{OU}}=1.0$), the fidelity remains very high (approximately $0.99$) and exhibits only minor variations with $\tau_c$, starting at $0.9942$ for $\tau_c=0.1$ and fluctuating around $0.988 - 0.991$ for $\tau_c \ge 0.5$. The corresponding Markovian model predicts a fidelity of $0.9951$.

As the noise strength increases, the impact of $\tau_c$ becomes more pronounced and non-monotonic. For $\sigma_{\text{OU}}=2.0$, the fidelity starts at $0.9753$ ($\tau_c=0.1$), dips to a local minimum of $0.9569$ at $\tau_c=1.0$, and then shows a slight recovery, fluctuating around $0.960 - 0.969$ for $\tau_c \ge 2.0$. The Markovian equivalent fidelity is $0.9797$.

At $\sigma_{\text{OU}}=3.0$, the fidelity begins at $0.9474$ ($\tau_c=0.1$), generally trends downwards with increasing $\tau_c$, reaching $0.9219$ at $\tau_c=10.0$, with intermediate local minima (e.g., $0.9293$ at $\tau_c=2.0$). The Markovian model predicts a higher fidelity of $0.9554$.

The non-Markovian characteristics are most evident at higher noise strengths. For $\sigma_{\text{OU}}=4.0$, using data from both standard and refined $\tau_c$ scans (see Fig.~\ref{fig:fid_vs_tau_c_full_sigma} and supplementary data if showing refined scan separately), the fidelity starts at $0.8983$ (main scan, $\tau_c=0.1$) or $0.8931$ (refined scan, $\tau_c=0.1$). It exhibits complex behavior, with notable dips observed. For instance, in the refined scan, minima were seen around $\tau_c=2.5$ ($0.8636$) and $\tau_c=8.0$ ($0.8387$). The main scan showed a significant dip to $0.8297$ at $\tau_c=0.5$. The Markovian fidelity for $\sigma_{\text{OU}}=4.0$ is substantially higher at $0.9197$.

For the strongest noise investigated, $\sigma_{\text{OU}}=5.0$, the fidelity starts at $0.8426$ (main scan, $\tau_c=0.1$) or $0.8471$ (refined scan, $\tau_c=0.1$). A sharp drop is observed, with a minimum fidelity of $0.7871$ occurring at $\tau_c=1.0$ (main scan). The refined scan showed minima such as $0.7819$ at $\tau_c=6.0$. For $\tau_c > 1.0$, the fidelity generally remains in the $0.78 - 0.83$ range, significantly below the Markovian prediction of $0.8761$.

These results demonstrate that for a given noise strength, there can exist intermediate correlation times where the fidelity is more severely degraded than in the limits of very short or potentially longer correlation times. Furthermore, for $\tau_c > 0.1 \Delta t_{\text{step}}$, the OU noise generally leads to lower fidelities than predicted by the equivalent Markovian model, with the discrepancy increasing with $\sigma_{\text{OU}}$.

\subsection{Impact of Noise Strength on Fidelity}
\begin{figure}[htbp]
  \centering
  \includegraphics[width=0.9\columnwidth]{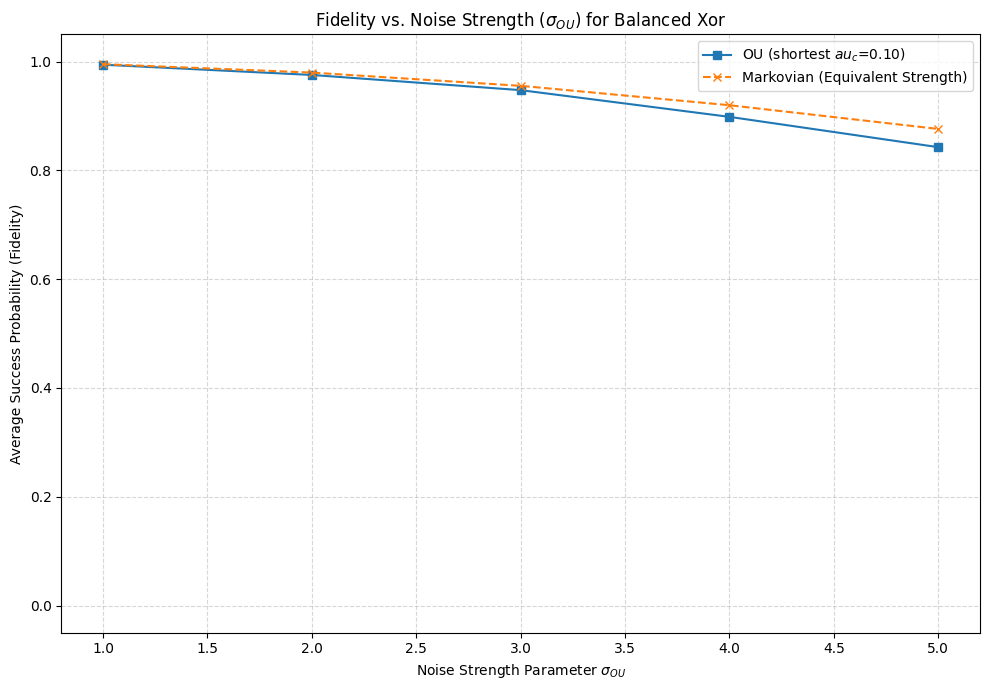}
  \caption{Fidelity of the 2-qubit Deutsch-Jozsa algorithm (balanced oracle) as a function of OU noise strength $\sigma_{\text{OU}}$. The solid line shows the OU model at the shortest correlation time simulated ($\tau_c=0.1 \Delta t_{\text{step}}$). The dashed line represents the fidelity under an equivalent Markovian dephasing model. Results averaged over 100 noise trajectories.}
  \label{fig:fid_vs_sigma}
\end{figure}

The overall effect of the noise strength parameter $\sigma_{\text{OU}}$ on the algorithm's fidelity is shown in Fig.~\ref{fig:fid_vs_sigma}.

As anticipated, the fidelity monotonically decreases as $\sigma_{\text{OU}}$ increases. The plot compares the OU noise model at its shortest simulated correlation time ($\tau_c=0.1 \Delta t_{\text{step}}$, solid line) with the Markovian dephasing model (dashed line), using data from simulations with 100 trajectories. There is a close agreement between these two models across the range of $\sigma_{\text{OU}}$ values. For instance, at $\sigma_{\text{OU}}=1.0$, the OU ($\tau_c=0.1$) fidelity is $0.9942$ compared to the Markovian fidelity of $0.9951$. At $\sigma_{\text{OU}}=5.0$, the OU ($\tau_c=0.1$) fidelity is $0.8426$, while the Markovian fidelity is $0.8761$.
While both models show the same qualitative trend, the OU noise at $\tau_c=0.1 \Delta t_{\text{step}}$ consistently yields a slightly lower fidelity than the matched Markovian model, with this difference becoming more noticeable at higher $\sigma_{\text{OU}}$. This suggests that even for very short correlation times, the OU noise model may capture effects not present in the idealized Markovian dephasing channel, or that the parameter matching condition (Eq.~\eqref{eq:lambda_pd_matching}) is an approximation that becomes less exact at higher noise strengths.

\section{Discussion}
\label{sec:discussion}

The simulation results presented in the previous section reveal a significant and complex interplay between temporally correlated dephasing noise and the fidelity of the 2-qubit Deutsch-Jozsa algorithm. Our findings highlight several key aspects regarding the impact of non-Markovian noise characteristics, particularly the noise strength ($\sigma_{\text{OU}}$) and correlation time ($\tau_c$), on a foundational quantum algorithm.

A primary observation is the non-monotonic dependence of the algorithm's fidelity on the noise correlation time $\tau_c$, especially at higher noise strengths (e.g., $\sigma_{\text{OU}} \ge 2.0$, as shown in Fig.~\ref{fig:fid_vs_tau_c_full_sigma}). Unlike simple Markovian dephasing, where the error probability per gate is typically assumed constant, the correlated nature of the Ornstein-Uhlenbeck (OU) noise leads to regimes where specific $\tau_c$ values are notably more detrimental to fidelity. For instance, with $\sigma_{\text{OU}}=5.0$, a distinct minimum in fidelity was observed around $\tau_c = 1.0 \Delta t_{\text{step}}$. This behavior can be attributed to the "memory" of the noise. When $\tau_c$ is comparable to the effective duration over which the qubit is susceptible to phase errors during critical stages of the algorithm (here, approximately $3 \Delta t_{\text{step}}$), the noise values at sequential error points are not independent. This can lead to an accumulation of phase errors that do not average out as effectively as in the very fast ($\tau_c \ll \Delta t_{\text{step}}$) or potentially very slow ($\tau_c \gg \text{total algorithm time}$) noise regimes. The observed dips suggest that noise fluctuating on timescales that align with the algorithm's operational segments can lead to a form of resonant error amplification or a persistent error bias through consecutive operations.

This study underscores a critical limitation of the standard Markovian approximation when characterizing the performance of quantum algorithms in realistic noise environments. Our results consistently show that while the OU noise model at very short correlation times ($\tau_c=0.1 \Delta t_{\text{step}}$) closely approximates the fidelity predicted by a Markovian dephasing model with an equivalently matched single-step error variance (Fig.~\ref{fig:fid_vs_sigma}), this approximation breaks down substantially for larger $\tau_c$. In most cases where $\tau_c > 0.1 \Delta t_{\text{step}}$ and the noise strength was significant, the Markovian model \textit{overestimated} the fidelity of the Deutsch-Jozsa algorithm (Fig.~\ref{fig:fid_vs_tau_c_full_sigma}). This implies that relying solely on Markovian assumptions for device benchmarking or error threshold estimations could lead to overly optimistic predictions of algorithmic success on near-term quantum hardware, where noise often exhibits temporal correlations \cite{Martinis2003_LowFreqNoise, Bylander2011_1fNoise}.

Interestingly, even in the quasi-Markovian limit ($\tau_c=0.1 \Delta t_{\text{step}}$), the OU noise model resulted in slightly lower fidelities than the matched ideal Markovian phase damping channel, particularly at higher noise strengths (Fig.~\ref{fig:fid_vs_sigma}). This minor discrepancy could arise from the fact that the OU noise, even with short $\tau_c$, still possesses some residual short-term memory not captured by the perfectly memoryless Markovian channel. Alternatively, it might reflect the limits of the parameter matching condition (Eq.~\eqref{eq:lambda_pd_matching}) which equates average effects, while the full distributions of errors might differ. The continuous nature of the phase errors $\phi_k$ drawn from a Gaussian distribution (due to OU noise) might also have a subtly different impact compared to the discrete error events often modeled in simpler Pauli or phase damping channels.

The implications of these findings are pertinent to the ongoing development of quantum computing technologies. The sensitivity to noise correlation times suggests that a more detailed characterization of the noise environment, beyond just its strength or average error rates (like $T_1, T_2$ times derived assuming Markovianity), is crucial. Understanding the full noise power spectral density, from which $\tau_c$ can often be inferred, may be necessary for accurate performance prediction and for designing effective error mitigation or correction strategies tailored to specific non-Markovian environments \cite{Klimov2018_NonMarkovianCharacterization, Szankowski2017_ReviewNonMarkovian}. While the Deutsch-Jozsa algorithm is relatively simple, its reliance on quantum interference for its speedup makes it a sensitive probe for phase errors, and the observed effects may well extrapolate to more complex algorithms that depend on coherent superpositions.

This study has several limitations. The noise was applied only to the query qubit, whereas in a real system, both qubits would be susceptible to noise, and crosstalk effects might introduce correlated noise between qubits. We considered a specific, albeit common, model for temporally correlated noise (OU process) and applied it at discrete points. Real quantum gates have finite durations, and noise might act continuously or have a more complex temporal profile during gate execution. Furthermore, the analysis was restricted to a 2-qubit system.

Future work could extend this investigation in several directions. Exploring other types of correlated noise, such as $1/f$ (flicker) noise which is prevalent in many solid-state qubit platforms, would be highly relevant. Investigating the impact of correlated noise on larger multi-qubit algorithms, including those with greater circuit depth, could reveal more complex cumulative effects. Furthermore, the development and testing of error mitigation techniques specifically designed to counteract temporally correlated dephasing, perhaps by adapting dynamical decoupling sequences based on measured noise correlation times \cite{Viola1999_DD, Biercuk2009_DD}, would be a valuable pursuit. Finally, efforts to bridge the gap between such simulation studies and experimental observations by using empirically determined noise parameters from real quantum devices would be essential for validating these theoretical insights.

In conclusion, our simulations demonstrate that temporally correlated dephasing noise can significantly impact the fidelity of the 2-qubit Deutsch-Jozsa algorithm in ways not captured by simple Markovian models. The fidelity exhibits a strong dependence on the noise correlation time, often leading to a greater performance degradation than might be expected from uncorrelated noise of similar strength, particularly when the correlation time is comparable to the operational timescales of the algorithm. These findings contribute to a deeper understanding of how realistic noise environments affect quantum computation and highlight the importance of considering non-Markovian characteristics in the design and analysis of quantum algorithms and devices.

\section{Conclusion}
\label{sec:conclusion}

In this study, we have investigated the impact of temporally correlated dephasing noise, modeled by the Ornstein-Uhlenbeck process, on the fidelity of the 2-qubit Deutsch-Jozsa algorithm. Our numerical simulations systematically explored a range of noise strengths ($\sigma_{\text{OU}}$) and correlation times ($\tau_c$), revealing important deviations from the behavior expected under simpler Markovian noise assumptions.

Our principal finding is that the fidelity of the Deutsch-Jozsa algorithm exhibits a significant and non-monotonic dependence on the noise correlation time $\tau_c$, particularly at higher noise intensities. We identified regimes where intermediate correlation times, comparable to or slightly exceeding the effective duration of the algorithm's noisy operations, lead to a more substantial degradation of fidelity than either very short or potentially very long correlation times. Specifically, for strong noise ($\sigma_{\text{OU}} = 5.0$), the minimum fidelity observed was $0.7871$ at $\tau_c=1.0 \Delta t_{\text{step}}$.

Furthermore, we demonstrated that a standard Markovian dephasing model, even when its strength is matched to the per-step error variance of the OU process, generally overestimates the algorithm's fidelity for $\tau_c > 0.1 \Delta t_{\text{step}}$. This highlights the limitations of memoryless noise models in accurately predicting quantum algorithm performance in environments where noise correlations are present. The extent of this overestimation by the Markovian model was found to increase with the overall noise strength.

These results underscore the critical importance of characterizing and accounting for the temporal structure of noise in quantum computing systems. The non-Markovian effects observed are not mere minor corrections but can lead to qualitatively different predictions of algorithmic success. For the continued development of robust quantum algorithms and effective error mitigation strategies on near-term quantum devices, a deeper understanding of realistic, correlated noise environments is indispensable. Our work contributes to this understanding by quantifying the impact of such correlations on a foundational quantum algorithm, paving the way for more accurate noise modeling and performance assessment in future quantum information processing tasks.



\end{document}